**Structural and magnetic properties in sputtered iron oxide epitaxial thin films**

**- Magnetite $Fe_3O_4$ and epsilon ferrite ε-$Fe_2O_3$ -**

Masato Watanabe

Research Institute for Electromagnetic Materials

9-5-1 Narita, Tomiya 981-3341 Japan

Phone: +81-22-341-6343

Fax: +81-22-347-3789

E-mail.: m_watanabe@denjiken.ne.jp, masato33@innovamateria.jp

**Abstract**

Epitaxial thin film fabrication of iron oxides including magnetite $Fe_3O_4$ and epsilon-ferrite ε-$Fe_2O_3$ with the potential for advancing electromagnetic devices has been investigated, which led to the first ever ε-ferrite epitaxial layer being synthesized in the conventional sputtering process. Concerning $Fe_3O_4$ (100) / MgO (100) films, a cube-on-cube epitaxial relationship and sharp rocking curves with FWHM of 50 - 350 arcsec were confirmed regardless of the small amount of Ge additions. Sputtering Ar gas pressure $P_{Ar}$ heavily influenced their magnetic and transport properties. High $P_{Ar}$ = 15 mTorr caused a high magnetization of 6.52 kG for the Ge added sample and the clear Verwey transition at 122 K for the non Ge addition case. Conversion electron Mössbauer spectroscopy (CEMS) measurements revealed that low $P_{Ar}$ < 10 mTorr causes Fe/O off-stoichiometry on the oxidizing side for the non Ge addition case and the reductive side for the Ge addition case, respectively. Regarding the ε-$Fe_2O_3$ (001) / $SrTiO_3$(111) epilayer synthesis, bilayer microstructure composed of an approximately 5nm thick initially grown ε-$Fe_2O_3$ (001) epilayer and subsequently grown α-$Fe_2O_3$ (001) epilayer was confirmed from cross-sectional TEM observations. The coexistence of magnetically hard and soft phases was confirmed from the magnetization measurements. As a possible application of the single nm thick ε-$Fe_2O_3$ layer, 4-resistive-state multiferroic tunnel junction (MFTJ) is considered.

## Introduction

Magnetic iron oxides, so-called "ferrites", have been utilized for various electromagnetic applications due to their versatile magnetic properties [1, 2] and large natural abundance of main constituent iron and oxygen, which meets recent social requirements for materials such as rare-metal-free ubiquity. Among the huge variety of ferrites, we focused on two iron oxides: magnetite $Fe_3O_4$ and epsilon ferrite $\varepsilon$-$Fe_2O_3$, both of which show unique magnetic and electronic functionalities, and attempted to fabricate their epitaxially grown thin films by conventional sputtering, which is an advantageous process for industrial device applications.

Magnetite of the first iron oxide is an ubiquitous magnetic material naturally found as the main component of iron sand [3]. Its crystal structure is an inverse spinel that is composed of $Fe^{3+}$ at tetrahedral A sites, $Fe^{2+}$ and $Fe^{3+}$ at octahedral B sites and oxygen sites. It has a room temperature saturation magnetization $4\pi M$ of 6.25 kG, which is the highest among iron oxides, and Curie temperature $T_C$ of 858 K [4]. Magnetite also shows characteristic electronic properties of half-metallicity [5-8] and large anomalous Hall resistivity $\rho_H$ ~10 - 40 μΩcm [9], which is comparable with Co based full Heusler compounds [10], leading to the possibility for various spintronic devices such as magnetic tunnel junctions (MTJ). Due to its high biocompatibility, magnetite's biomedical applications such as hyperthermia and drug delivery system (DDS) have also been pursued [11]. To date, our research group conducted research on small-amount element additions in polycrystalline magnetite films and found that the additions of some elements up to several percent, especially Ge, raised their thermal stability and caused an increase in magnetization [12, 13]. Since the effects of such element additions in magnetite epifilms have not yet been confirmed, we studied the structural and magnetic effects on small-amount Ge addition to magnetite epifilms in this research.

Epsilon ferrite of the second iron oxide is an emerging hard magnetic material, which has an orthorhombic crystal structure with four iron sites for which magnetic moments are ferrimagnetically aligned [14]. From its high anisotropy field $H_A$ that comes from the high magnetic anisotropy constants of $K_a = 7.7 \times 10^6$ erg/cc along the a axis and $K_b = 1.2 \times 10^6$ erg/cc along the b axis [15] and low magnetization value $4\pi M_S = 1.26$ kG (= 100 emu/cm$^3$) [16], ε-ferrite shows high resonance frequencies $f_r$ of 35-222 GHz [17, 18] and high coercivities over 20 kOe [17, 19], leading to the possibility of milli-wave absorption device and magnetic recording media applications. It also shows a rare room-temperature multiferroicity that means simultaneous manifestation of ferromagnetism and ferroelectricity [16], which may have memory or sensor applications as indicated in the last part of this paper. Its synthesis is difficult due to the characteristics of the metastable phase. Two preparation methods available have been proposed for metastable ε-ferrite so far. One is based on Gibb's energy minimum of ε-ferrite nanoparticles at the particle diameters of 8-30 nm [17], leading to the availability of chemical syntheses of ε-ferrite nanoparticles. The other is based on the constraint of crystal lattices accompanied by epigrowth on certain single-crystalline substrates, which has mainly been conducted for ε-ferrite epifilms by pulsed laser deposition (PLD). A few single-crystalline substrates including $SrTiO_3$(111) [16, 20], YSZ(100) [21] and GaN(0001) [22, 23] have been reported for the syntheses of ε-ferrite epifilms by PLD. However, epifilm synthesis by conventional sputtering, which is important in industrial applications, has to the best of my knowledge not yet been reported. Therefore, we attempted ε-ferrite epifilm synthesis by RF magnetron sputtering.

**Experimental**

Sample preparations for both magnetite and ε-ferrite epifilms were done by RF magnetron sputtering. The sputtering target for the magnetite epilayer was a magnetite composite target on which Ge chips were regularly-arrayed and that for the ε-ferrite epilayer was an $\alpha$-$Fe_2O_3$ target, respectively. Single-crystalline substrates for epigrowth were polished MgO(100) (Tateho Chemical Industries) for magnetite epilayers and polished $SrTiO_3$(111) (Furuuchi Chemical Corp.) for ε-ferrite epilayers, both of which were heated up to 800°C during deposition after evacuating at less than the vacuum degree of 2 x $10^{-7}$ Torr. Sputtering gas species were pure Ar gas for magnetite epigrowth and Ar + $O_2$ mixed gas with the $O_2$ / (Ar +$O_2$) flow ratio ranging from 25% to 50% for ε-ferrite epigrowth, respectively. Sputtering gas pressures, which strongly affect film characteristics, were varied from 2 to 15 mTorr for magnetite epigrowth, and from 1 to 5 mTorr for ε-ferrite epigrowth, respectively. The film thicknesses for magnetite epilayers ranged from 0.481 μm to 0.548 μm, which were measured by DekTak 150 (Bruker). Ge compositions in magnetite epilayers were evaluated by wavelength dispersive X-ray spectroscopy WDS (INCA WAVE, Oxford Instruments). Structural analyses were conducted using a high-precision X-ray diffractometer with a CuKα radiation, LYNXEYE 1-dimensional detector and Ge(220) monochromator (D8 DISCOVER, Bruker). The film thicknesses of $\alpha/\varepsilon$-$Fe_2O_3$ bilayers were evaluated by X-ray reflectometry. Magnetization measurements were performed by a superconducting quantum interference device SQUID (MPMS3, Quantum Design) and vibrating sample magnetometer VSM (Tamakawa) with maximum applied fields of 70 kOe and 15 kOe, respectively. Diamagnetic components that come from substrates and sample holders were subtracted from the raw magnetization data. Conversion electron Mössbauer spectroscopy (CEMS) measurement was performed with a $^{57}Co$ radiation embedded in a Rh matrix, constant-acceleration spectrometer and He-1% $(CH_3)_3CH$ gas-flow counter. Velocity correction was conducted using α-Fe. No $^{57}Fe$ was doped in the samples for CEMS. All the measurements were performed at room temperature. Resistivity measurements were performed by the 4-point method using PPMS (Quantum Design) in the temperature range of 10 - 300K. Cross-sectional high-angle annular dark-field scanning transmission electron microscopy (HAADF-STEM) observation (Z-contrast image) and nano beam electron diffraction (NBD) for the α/ε-$Fe_2O_3$ bilayer were conducted by JEM-ARM200F (200 kV, JEOL).

**Results and discussions for $Fe_3O_4$ epigrowth**

A typical θ/2θ scan XRD profile for the 0.6 at%Ge-$Fe_3O_4$ / MgO(100) epifilm is shown in Fig. 1. Sharp $Fe_3O_4$ (400) and $Fe_3O_4$ (800) peaks were clearly observed close to the MgO(200) and MgO(400) peaks and no other phase was confirmed, which shows c-axis alignment of the magnetite phase in the direction normal to the film plane. In order to study the effect of sputtering gas pressure $P_{Ar}$ on crystallinity of the magnetite epilayers, full-widths at half maximum (FWHM) of $Fe_3O_4$ (400) rocking curves with varying $P_{Ar}$ are shown in Fig. 2 in comparison with FWHM of MgO(200). FWHM of the magnetite epilayers was relatively low ranging from 50 to 350 arcsec, which is the same order of magnitude as those of compound semiconductor epifilms reported so far. The FWHM of magnetite epilayers for $P_{Ar} \leq 7$ mTorr was lower than those for $P_{Ar} \geq 10$ mTorr, indicating that magnetite epilayers with lower $P_{Ar} \leq 7$ mTorr have higher crystallinity than the higher $P_{Ar} \geq 10$ mTorr cases, and shows almost the same crystallinity as those of MgO(200). Conventional θ/2θ scan XRD gives only structural information along the out-of-plane direction, and therefore we conducted $Fe_3O_4$(311) φ scan XRD that contains its in-plane information, as shown in Figs.3 (a) and (b). Different from $Fe_3O_4$ (400), $Fe_3O_4$(311) is appropriate for φ scan XRD since it is sufficiently separated from the MgO substrate spots in its reciprocal lattice space. Figure 3 (c) shows a simulated $Fe_3O_4$(311) pole figure with the φ = 0° direction along $Fe_3O_4$ [100], of which the inner four spots and outer eight spots correspond to the four peaks in Fig. 3 (a) and the eight peaks in Fig.3 (b), respectively. Reproduction of the simulated pole figure is considered to show the following cube-on-cube orientation relationships: $Fe_3O_4$[100] // MgO[100], $Fe_3O_4$[010] // MgO[010] and $Fe_3O_4$[001] // MgO[001].

We have studied room-temperature magnetic properties for the $Fe_3O_4$(100) / MgO(100) epifilms varying with Ar sputter gas pressure $P_{Ar}$. Magnetization curves for $P_{Ar}$ = 15 mTorr and $P_{Ar}$ = 2 mTorr with the maximum field of 70 kOe are shown in Figs. 4 (a) and (b). Out-of-plane magnetizations for both $P_{Ar}$ = 15 mTorr and 2 mTorr exceed the in-plane magnetizations at higher fields over 10 kOe, which suggests the existence of some perpendicular magnetic anisotropy. This large perpendicular anisotropy has been reported for magnetite epifilms and polycrystalline films [24, 25] so far and its possible origin is attributed to out-of-plane moment alignment around antiphase boundaries (APB), which was confirmed by off-axis electron holography [26]. Figure 4 (c) summarizes out-of-plane magnetizations 4πM with the maximum applied fields of 15 kOe and 70 kOe for the Ge-doped and non-doped magnetite epifilms as a function of $P_{Ar}$. For both the Ge-doped and non-doped cases, magnetizations 4πM increase with the increase in $P_{Ar}$. Moreover, the out-of-plane 4πM for $P_{Ar}$ = 15 mTorr is 6.52 kG, which exceeds the 6.25 kG for single crystal magnetite, at the applied field of 70 kOe and still does not saturate at 70 kOe as shown in Figs.4 (a) and (b). Magnetization enhancement over 12 kG, which is much larger than the bulk value of 6.25 kG, has been reported for 3nm thick magnetite epifilm, the possible origin of which has been attributed to spin flip from ferrimagnetic moment alignment to ferromagnetic type [4]. The same type of enhancement might occur partially for the epifilms as well. In this research, we grew magnetite epifilms at a high substrate temperature of 800°C and magnesium atoms easily diffuse into the magnetite layer [27, 28], leading to the possible formation of an interdiffused Mg-$Fe_3O_4$ layer. Thus, we conducted cross-sectional TEM observation and composition analysis along the out-of-plane direction, which confirmed the existence of a ~100 nm thick Mg-$Fe_3O_4$ interdiffused layer in the magnetite epifilm with a total thickness of ~500 nm. Mg ferrite $MgFe_2O_4$ has 4πM of 1.149 kG [29], which is much lower than that of the 6.25 kG for magnetite, and so the interdiffused layer is considered to cause a magnetization reduction. Since a slight magnetization enhancement was observed for the magnetite epilayer with $P_{Ar}$ = 15 mTorr as described above despite the existence of the interdiffused layer, the magnetization enhancement at the undiffused part of magnetite epilayer is considered to be large enough to compensate the magnetization reduction at the Mg interdiffused part.

In order to study the Verwey transition, which is sensitive to the film quality and stoichiometry [30], we performed resistance measurements of the (a) Ge-doped and (b) non-doped magnetite epifilms with $P_{Ar}$ = 10 and 15 mTorr as a function of the inverse number of temperature 1000/T as shown in Fig. 5. The non-doped magnetite epifilms show the clear resistivity change accompanied with the Verwey transition. The transition temperature $T_V$ for $P_{Ar}$ = 15 mTorr is 122 K, almost the same as the previously reported value of bulk magnetite. On the contrary, resistivity changes for the Ge-doped samples are diminished and their $T_V$'s decrease until < 100 K. Since the Ge addition also increases the resistivity, the added Ge might be incorporated into B sites with two types of valency and inhibit the hopping conduction. Regardless of the Ge concentrations, we could not observe clear resistivity changes of the Verwey transition for the samples with $P_{Ar}$ < 10 mTorr, the magnetizations of which are low as shown in Fig. 4(c).

For further investigation of magnetism in the magnetite epifilms, we performed CEMS measurements. CEMS spectra for the Ge-added magnetite epifilms are shown in Fig.6 (a) $P_{Ar}$ = 15 mTorr (0.4 at%Ge) and (b) $P_{Ar}$ = 2 mTorr (1.6 at%Ge). The observed spectra can be decomposed into two sharp ferromagnetic sextets that come from tetrahedral A ($Fe^{3+}$) and octahedral B ($Fe^{2.5+}$) sites. Hyperfine structure parameters, $B_{Hf}$, $\delta_{MB}$ integral intensities for A and B sites and sextet intensity ratios x for the Ge-added magnetite epifilms are summarized in Table 1 in comparison with those of previously reported sputtered magnetite epifilms and natural bulk magnetite [31, 32]. The parameters for the $P_{Ar}$ = 15 mTorr sample are in good agreement with those of the previously reported film and bulk magnetite. $B_{Hf}$ and integral intensities for the $P_{Ar}$ = 2 mTorr sample, which has lower magnetization 4πM than the $P_{Ar}$ = 15 mTorr sample, deviate from the previously reported values. Sextet intensity ratio x contains information about moment direction according to the following Eq.: $x = 4\sin 2\theta/(1+\cos 2\theta)$, where $\theta$ stands for the angle between $^{57}Fe$ magnetic moment and the incident direction of γ-ray for CEMS. States for x = 0 and x = 4 mean moment alignments along the out-of-plane and the in-plane directions, respectively. State for x = 2 means randomly-oriented or <111> oriented moments. From x < 2 for both the $P_{Ar}$ = 15 mTorr and 2 mTorr samples, the samples show some perpendicular magnetic anisotropy, which coincides with the feature of magnetization curves in Fig. 4 (a) and (b). Integral intensity ratios for A and B sites in CEMS contain information about iron vacancy in magnetite. Vacancy parameter $\delta$ in $Fe_{3-\delta}O_4$ has the following relation with the ratio of integral intensity for A site to that for B site $\beta$: $\beta = Integral\ Int.(A\ site)/Integral\ Int.(B\ site) = (1+5\delta)/(2-6\delta)$ [33]. $\delta$ = 0 and $\delta$ = 1/3 correspond to stoichiometric magnetite and stoichiometric γ-$Fe_2O_3$, respectively. If we take account of the effect of the recoilless fraction, a slight correction to the oxidizing side from $\beta$ = 0.5 ($\delta$ = 0) to $\beta$ = 0.52 ($\delta$ = 4.9 x $10^{-3}$) is required for stoichiometric magnetite [34]. Figures 7 (a) and (b) show $\delta$ and 4πM at 15 kOe for Ge-doped and non-doped MNT epifilms as a function of $P_{Ar}$. With the increase in $P_{Ar}$, δ's approach zero for both the Ge-doped and non-doped samples, which corresponds to the increase in 4πM. Therefore, the 4πM reduction with the decrease in $P_{Ar}$ is considered to be due to the composition deviation from the stoichiometry of magnetite. The vanishing of resistivity change at the Verwey transition with the decrease in $P_{Ar}$ described in the previous part of Fig. 5 is considered to be the same as well, and so Fe/O composition deviation is considered to significantly influence the magnetic and transport properties of magnetite.

**Results and discussions for α/ε-$Fe_2O_3$ bilayer epigrowth**

Research on epigrowth of ε-ferrite so far has been mainly focused on the PLD process and the conditions necessary for its epigrowth have been reported to be a high substrate temperature and high oxygen content atmosphere during deposition [16, 20-23]. Thus, we conducted sputtered epigrowth deposition at the high substrate temperature of 800°C with a varying oxygen flow rate ratio $Q(O_2)/Q(Ar+O_2)$. For the no oxygen atmospheric condition, the $Fe_3O_4(111)$ epilayer was grown despite the use of the $Fe_2O_3$ sputter target and found that ε-ferrite epilayers can be grown in a high oxygen content atmosphere such as the PLD cases as shown below. This study is the first to report of ε-ferrite epilayer growth by conventional sputtering process. XRD profiles for the $Fe_2O_3$ epifilms with $Q(O_2)/Q(Ar+O_2)$ = 25 % and different sputter gas pressures are shown in Figs.8 (a), (b) and (c). A c-axis oriented ε-ferrite epilayer was confirmed only for the lowest gas pressure of a 1 mTorr sample and the higher gas pressure conditions of 2 and 5 mTorr create α-phase epilayers. Low sputter gas pressure causes long mean free paths of sputter particles, leading to high kinetic energies, less collisions and low sputter particle incorporation into the films, which may influence the different phase formation in the epilayers. The prepared film samples on 10 mm square $SrTiO_3(111)$ substrates were composed of two different parts of the dark-colored and light-colored areas as shown in the optical images of Fig. 9. From XRD with narrowed X-ray beams, the dark-colored and light-colored areas are composed of the c-axis oriented α-phase and ε-phase, respectively. X-ray reflectometry with collimated beams shows that the thicknesses of the dark-colored and light-colored areas are 29 nm and 4 nm, respectively.

In order to clarify the microstructure of the epilayer, we conducted cross-sectional TEM observations on the dark-colored area of the epifilm with $Q(O_2)/Q(Ar+O_2)$ = 40 % and sputter gas pressure $P(Ar+O_2)$= 1 mTorr. Figure 10 shows cross-sectional TEM observation and NBD in the epifilm. It was revealed that the epifilm was composed of two phase construction where one is ε phase of the initial epilayer on the substrate with ~ 5 nm thick and the other is α-phase of the capped epilayer. From the NBD of Fig. 10 (b), both the ε- and α-phase epilayers were aligned with [001] along the out-of-plane direction and [-210] along the in-plane direction. The initial ε-phase epilayer is continuous in this image, however, discontinuous areas were also found to exist from larger area TEM observations. Since a clear lattice image was confirmed in Fig.10, we also conducted HAADF-STEM observation with higher magnification as shown in Fig.

11. Within the scope of our TEM observation, no crystalline boundaries were confirmed in the ε-phase epilayer though some disorder of atomic arrangement was observed, which is different from the case of the previously reported ε-phase epilayer prepared by PLD. It has been reported that PLD can create a thicker c-axis oriented ε-phase epilayer up to about 100nm thick and the PLD epilayer is composed of nano-sized crystal domains with the in-plane crystal orientation of six-fold symmetry [20]. ε-ferrite has been mainly studied within the form of nanoparticles and its phase formation is explained by free energy minimum at the diameters of 8-30nm [17]. However, other forms of epitaxially grown ε-ferrite have also been reported such as nanowire or nanobelt several micrometers in length by PLD [35, 36] and peculiar dendritic microstructures of mullite on the scale of several micrometers in Japanese stoneware, Bizen-yaki [37, 38]. Therefore, further study is considered to be necessary regarding the generation mechanism of ε-ferrite.

Room-temperature magnetic properties of the two phase α/ε-$Fe_2O_3$ epilayer with different total thicknesses were investigated as shown in Fig. 11. The hysteresis curves show that the sample was composed of decoupled magnetically soft and hard phases. The hard phase comes from original hard ε-ferrite and the soft phase comes from ε-ferrite in which magnetic anisotropy was diminished by some imperfections of crystallinity since no other magnetic phase could be confirmed. The magnetization values obtained by assuming 5 nm thick ε-ferrite epilayers had mostly good agreement with the reported value of 1.26 kG in ε-ferrite by the same order of magnitude, which is consistent with the above TEM observation results.

As a possible application of such single nm-scale ε-ferrite epilayer, it may be applied to the multiferroic barrier in multiferroic tunnel junction (MFTJ). Spin filter MFTJ with half-metallic and multiferroic perovskite oxides was firstly reported by a French group in 2007 as a prototype of the 4-resistive-state memory that utilizes both tunnel magnetoresistance (TMR) and tunnel electroresistance (TER) effects [39, 40]. 4-resistive states in MFTJ correspond to the configurations of magnetic moment M and electric polarization P as shown in Fig. 12. Because the high-temperature phase of magnetite and ε-ferrite shows half-metallicity and multiferroicity at room temperature, respectively, we anticipated that all ferrite MFTJ would work at room-temperature if it were composed of a magnetite electrode and ε-ferrite barrier as shown in Fig. 12. In order to secure the independent switching function of 4-state memory, ME coupling between M and P in the multiferroic barrier should not be strong and so the clarification of magnitude of ME coupling in ε-ferrite is required. A study on optical ME coupling in ε-ferrite utilizing its second harmonic generation SHG effect showed that it has a strong ME coupling from the magnetic component of SHG [15], thus it might be necessary to control the magnitude of coupling by measures such as some element additions in order to realize 4-resistive-state magnetite / ε-ferrite MFTJ.

**Summary**

Epitaxial growth of magnetite and ε-ferrite was conducted by conventional RF magnetron sputtering on MgO(100) and $SrTiO_3$(111), respectively. $Fe_3O_4$ / MgO(100) epifilms have a cube-on-cube epitaxial relationship and a high crystallinity was confirmed from the sharp rocking curves with FWHM of 50 - 350 arcsec for both the Ge-added and non Ge addition cases. Sputtering Ar gas pressure $P_{Ar}$ significantly influenced their magnetic and transport properties. The magnetite epifilm for $P_{Ar}$ = 15 mTorr had a high magnetization of 6.52 kG for the Ge added case and the clear Verwey transition at 122 K for the non Ge addition case. CEMS measurements revealed that low $P_{Ar} < 10$ mTorr causes Fe/O off-stoichiometry on the oxidizing side for the non Ge addition case and the reductive side for the Ge addition case, respectively.

Regarding ε-$Fe_2O_3$ epilayer synthesis, bilayer microstructure composed of an approximately 5nm thick ε-$Fe_2O_3$ (001) epilayer initially grown on $SrTiO_3$(111) and a subsequently grown α-$Fe_2O_3$ (001) epilayer was confirmed from cross-sectional TEM observations. The coexistence of magnetically hard and soft phases was confirmed from the magnetization measurements. As a possible application of the single nm thick ε-$Fe_2O_3$ layer, 4-resistive-state multiferroic tunnel junction (MFTJ) could be considered.

**Acknowledgements**

This article was supported by JSPS KAKENHI Grant number 17K06806, "Magnetic anisotropy control by coherency strain in element-added magnetite thin films" and partially presented at 2nd International Conference on Magnetism and Magnetic Materials, Budapest, Hungary in September 24-26, 2018 (Allied Academies). I thank Dr. S. Abe, President K. Arai of Research Institute for Electromagnetic Materials, Prof. S. Sugimoto and Prof. K. Takanashi of Tohoku Univ. for their beneficial discussions, and also thank Dr. M. Ikeda of Quantum Design Japan, Ms. A. Tsutsui of Foundation for Promotion of Material Science and Technology of Japan (MST), Dr. T. Segi of KOBELCO Research Institute and Mr. H. Sato of Research Institute for Electromagnetic Materials for their assistance with and discussions on the characterizations.

Table 1. Hyperfine structure parameters obtained from CEMS for $Fe_3O_4$/MgO(100) epifilms with $P_{Ar}$ = 15 mTorr (0.4 at%Ge) and $P_{Ar}$ = 2 mTorr (1.6 at%Ge). Previously reported data for sputtered $Fe_3O_4$ epifilms [31] and natural bulk $Fe_3O_4$ [32] are also shown for comparison.

| | $^{57}$Fe site | Hyperfine field $B_{hf}$ (kOe) | Isomer shift $\delta_{MB}$ (mm/s) | Integral Intensity for A & B sites (%) | Intensity ratio x (3:x:1:1:x:3) | 4πM (kG) |
|---|---|---|---|---|---|---|
| **$Fe_3O_4$ - 0.4at%Ge (15 mTorr)** | A ($Fe^{3+}$) | 492 | 0.28 | 35.0 % | 1.56 | 5.8 |
| | B ($Fe^{2.5+}$) | 461 | 0.67 | 65.0 % | | |
| **$Fe_3O_4$ - 1.6at%Ge (2 mTorr)** | A ($Fe^{3+}$) | 485 | 0.28 | 27.7 % | 1.63 | 4.5 |
| | B ($Fe^{2.5+}$) | 457 | 0.64 | 72.3 % | | |
| **Sputtered $Fe_3O_4$ film** | A ($Fe^{3+}$) | 491 ± 2 | 0.27 ± 0.01 | 33.3 ± 2% | 0.71 - 1.60 | 4.9 - 6.0 |
| | B ($Fe^{2.5+}$) | 462 ± 2 | 0.67 ± 0.01 | 66.6 ± 2% | | |
| **Natural Bulk $Fe_3O_4$** | A ($Fe^{3+}$) | 493 ± 1 | 0.27 ± 0.03 | 33.8 ± <2% | 2 | --- |
| | B ($Fe^{2.5+}$) | 460 ± 1 | 0.67 ± 0.03 | 66.2 ± <2% | | |

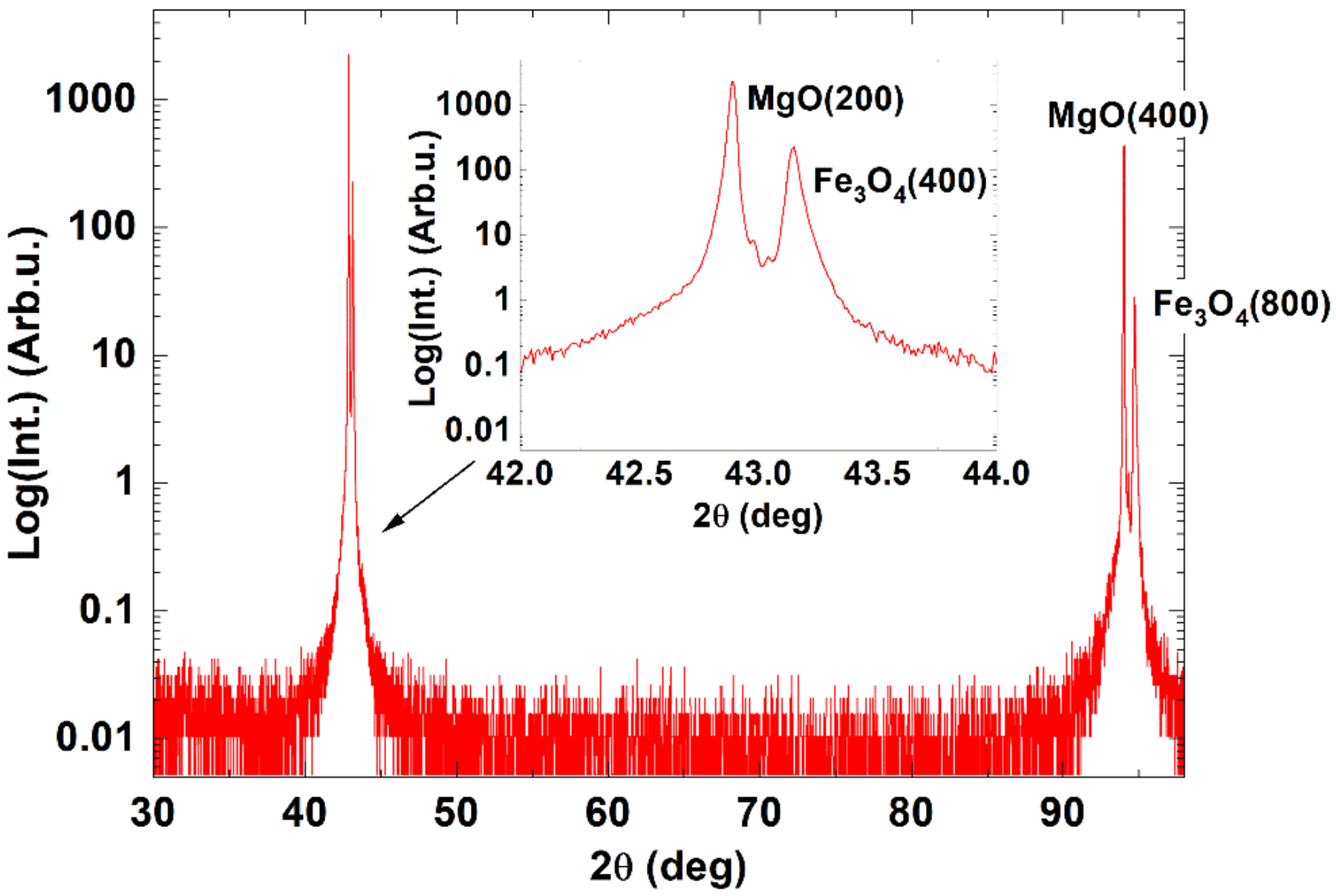

Fig. 1. θ/2θ scan XRD profiles for the 0.6 at%Ge added $Fe_3O_4$ epifilms grown on MgO (100). Magnified profile around MgO(200) and $Fe_3O_4$(400) peaks is shown in the inset.

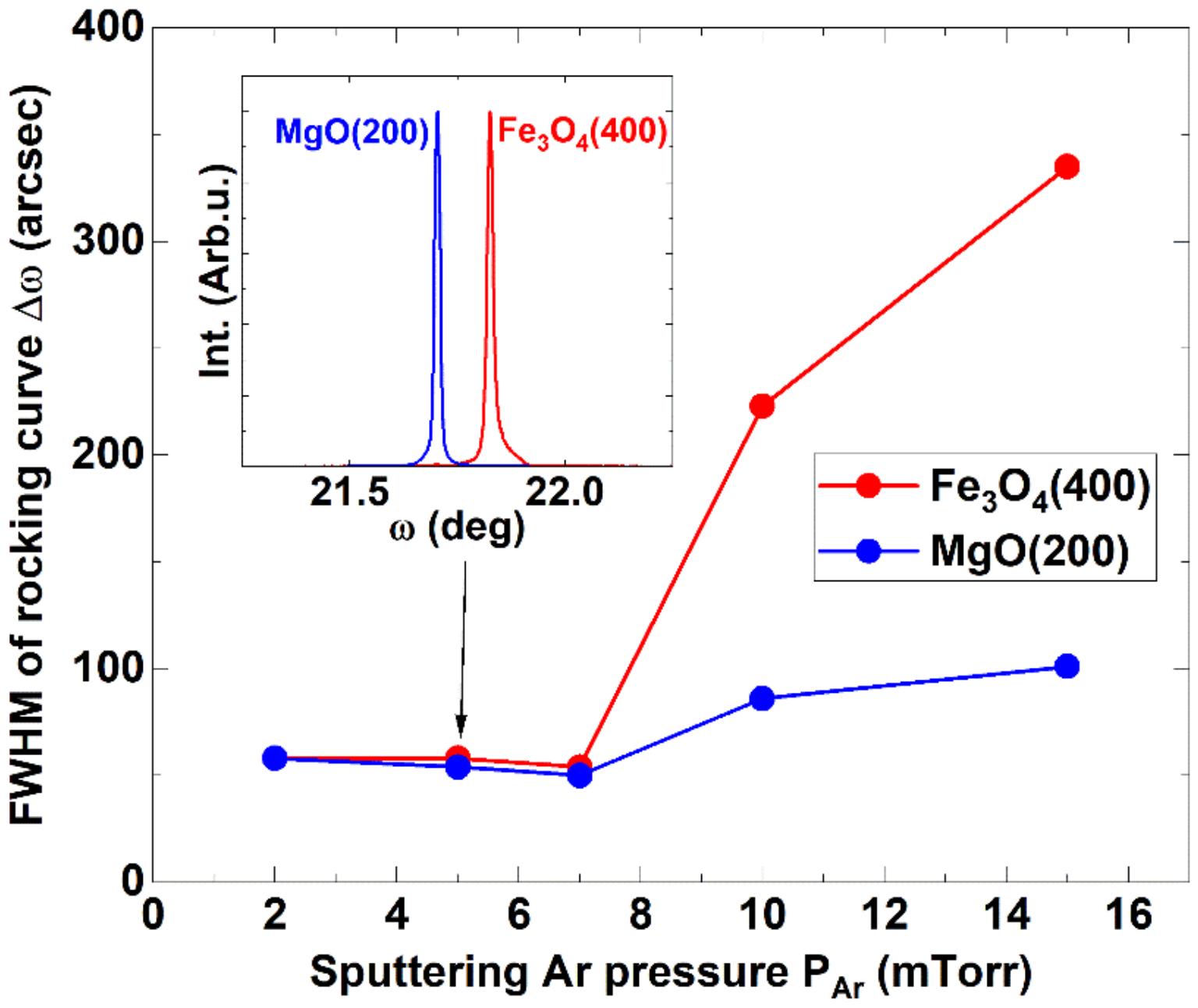

Fig. 2. Full widths at half maximum (FWHM) of $Fe_3O_4$(400) and MgO(200) rocking curves as a function of sputtering Ar pressure $P_{Ar}$ for the $Fe_3O_4$ / MgO(100) epifilms. The inset shows MgO(200) and $Fe_3O_4$(400) rocking curve profiles for the epifilm at $P_{Ar}$ = 5mTorr.

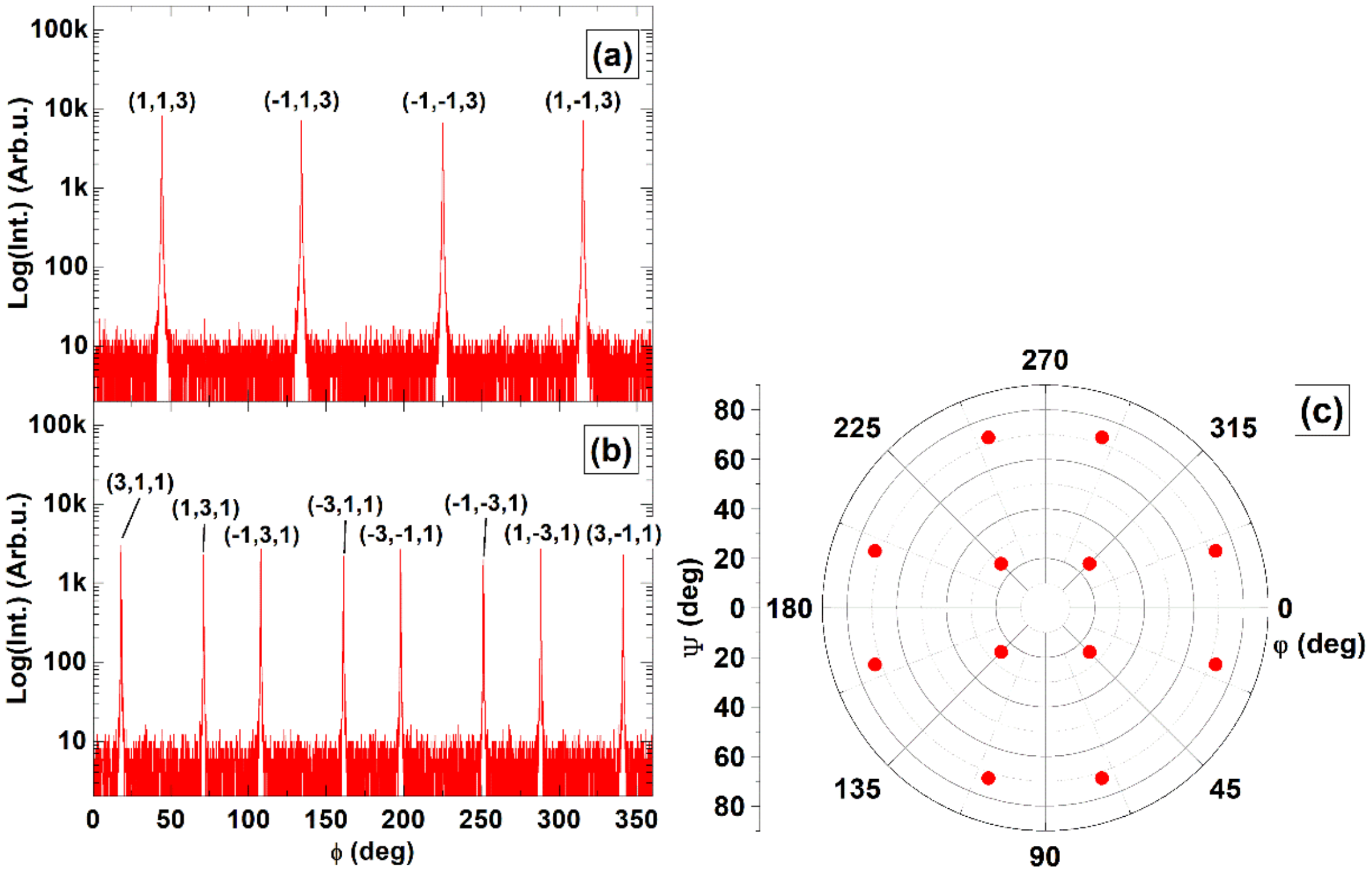

Fig. 3. $Fe_3O_4$ (311) φ scan profiles for the $Fe_3O_4$/MgO(100) epifilms at (a) ψ = 25.2393 deg and (b) ψ = 72.4512 deg, which correspond to the four inner and eight outer spots in (c) simulated $Fe_3O_4$ (311) pole figure.

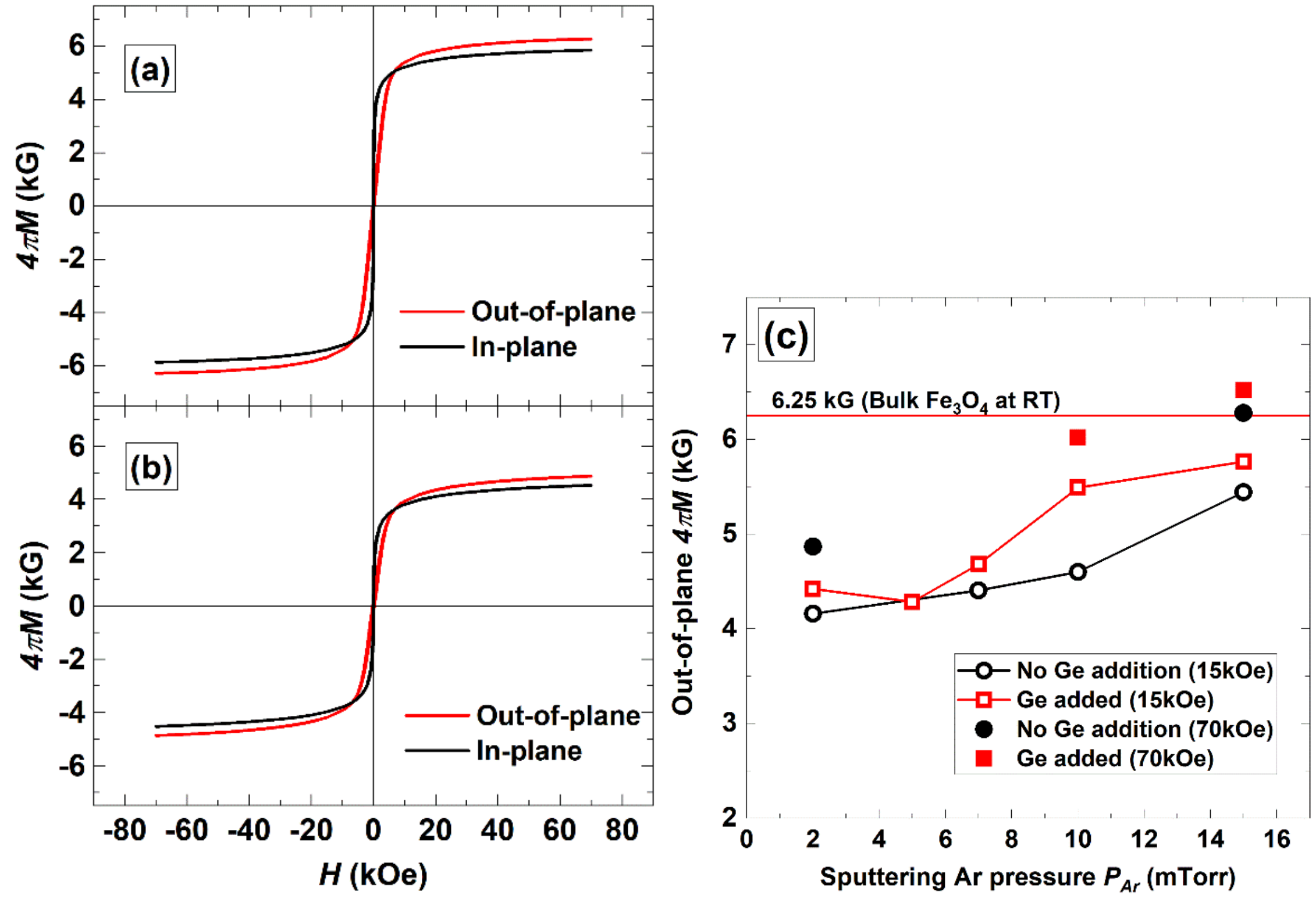

Fig. 4. Room-temperature magnetization hysteresis curves for the $Fe_3O_4$(100)/MgO(100) epifilms with non Ge addition at (a) $P_{Ar}$ = 15 mTorr and (b) $P_{Ar}$ = 2 mTorr. The red and black lines stand for out-of-plane and in-plane curves, respectively. (c) Out-of-plane magnetization 4πM at 15 kOe and 70 kOe as a function of sputtering Ar gas pressure $P_{Ar}$. Red blank squares and black blank circles stand for Ge added and non Ge addition cases at 15 kOe. Solid marks are 4πM at 70 kOe.

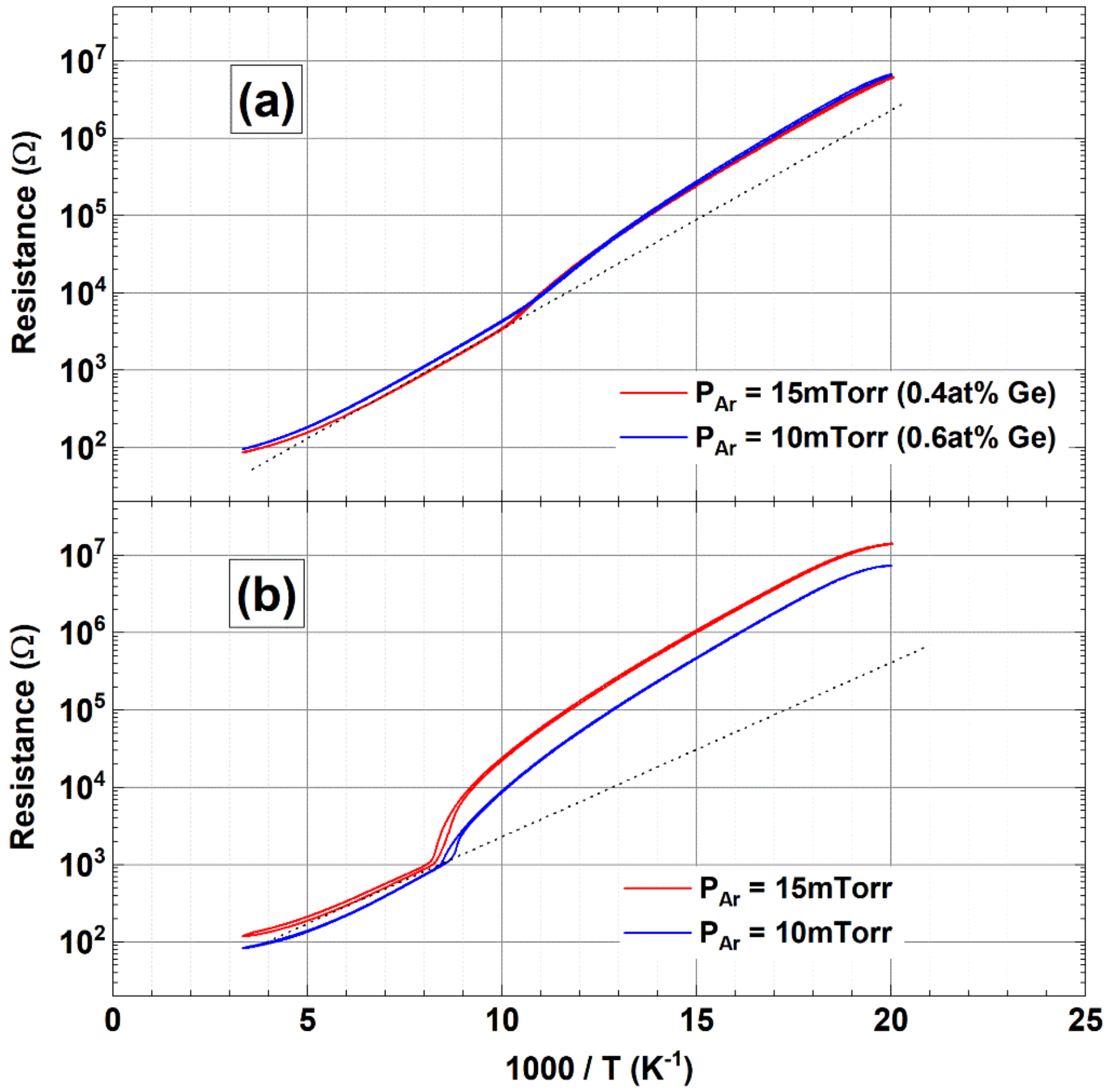

Fig. 5. Resistance as a function of 1000/T for (a) the Ge-added and (b) non Ge added $Fe_3O_4$(100)/MgO(100) epifilms with sputtering Ar gas pressure $P_{Ar}$ = 15 and 10 mTorr.

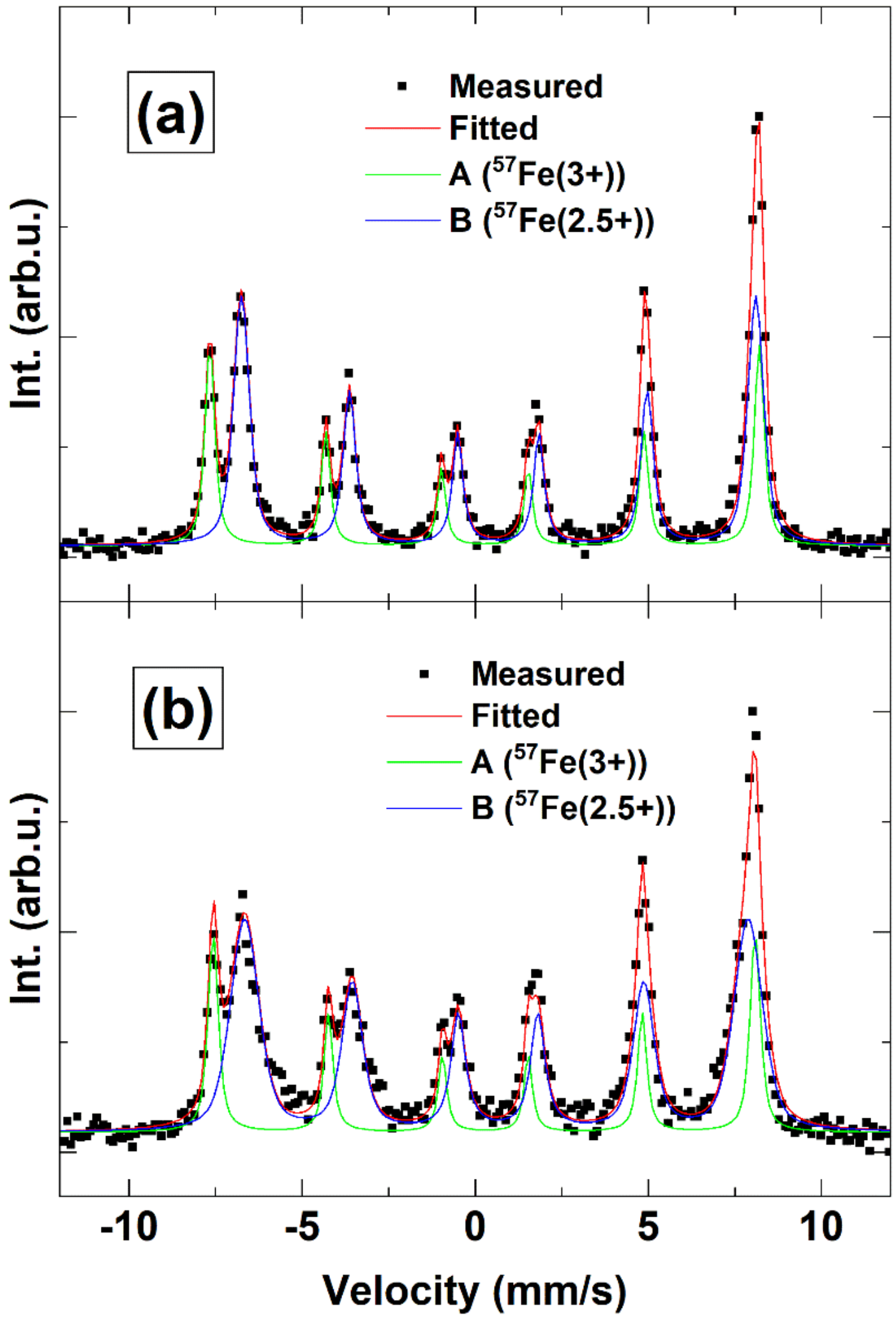

Fig.6. Converson electron Mossbauer spectroscopy (CEMS) spectra for the $Fe_3O_4$(100)/MgO(100) epifilms with (a) $P_{Ar}$ = 15 mTorr (0.4 at%Ge) and (b) $P_{Ar}$ = 2 mTorr (1.6 at%Ge). Measured CEMS data (black dots and fitted red lines) are decomposed into A site (green lines) and B site (blue lines) components.

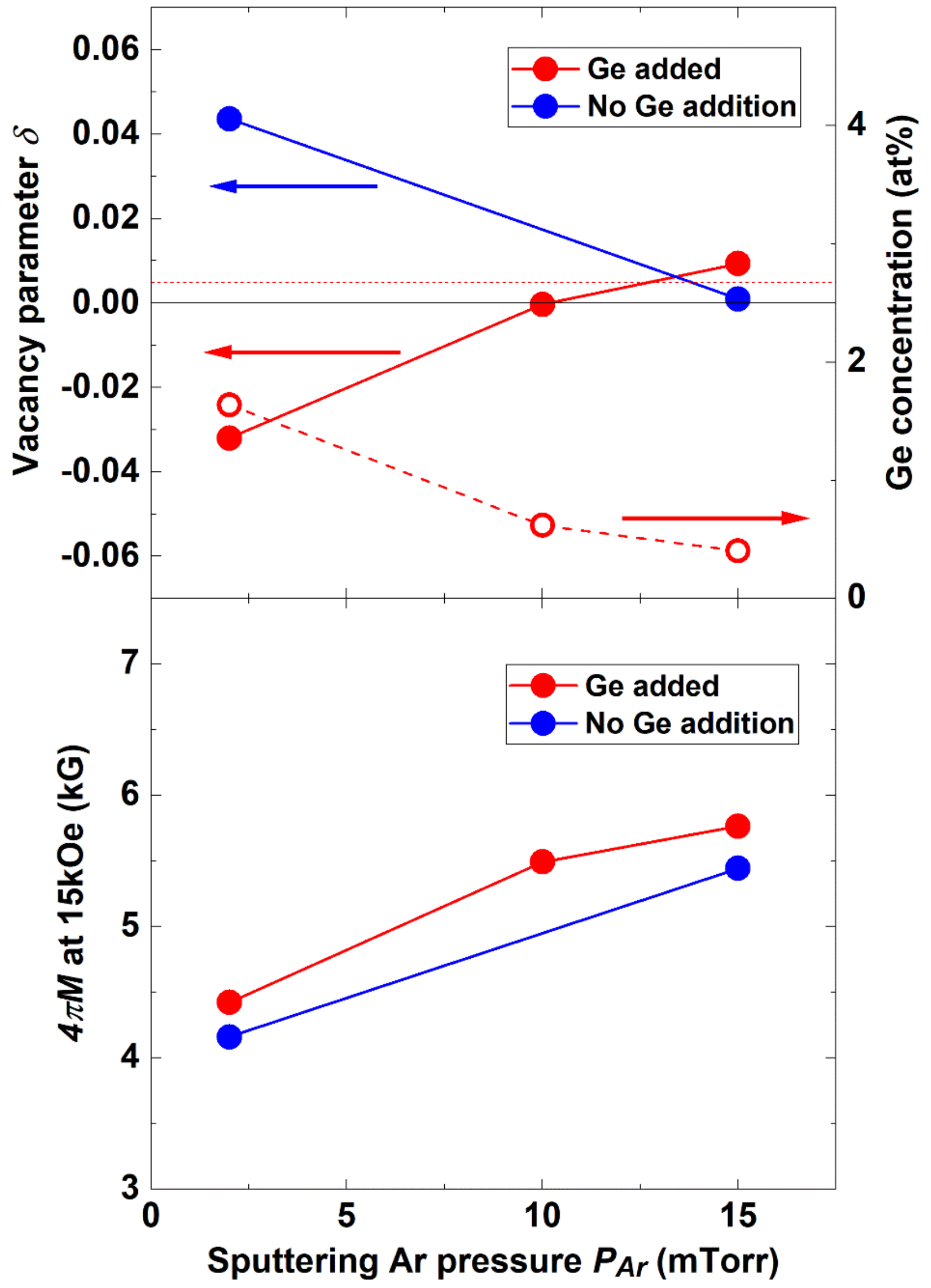

Fig. 7. (a) Vacancy parameter δ (solid red circles for Ge added and blue circles for non Ge addition cases) and Ge concentration (blank red circles), and (b) room-temperature magnetization 4πM at 15 kOe for the $Fe_3O_4$(100)/MgO(100) epifilms as a function of sputtering Ar pressure $P_{Ar}$. The ideal $\delta = +4.9 \times 10^{-3}$ for stoichiometric $Fe_3O_4$ taking its recoilless fraction into account is indicated with a red dotted line.

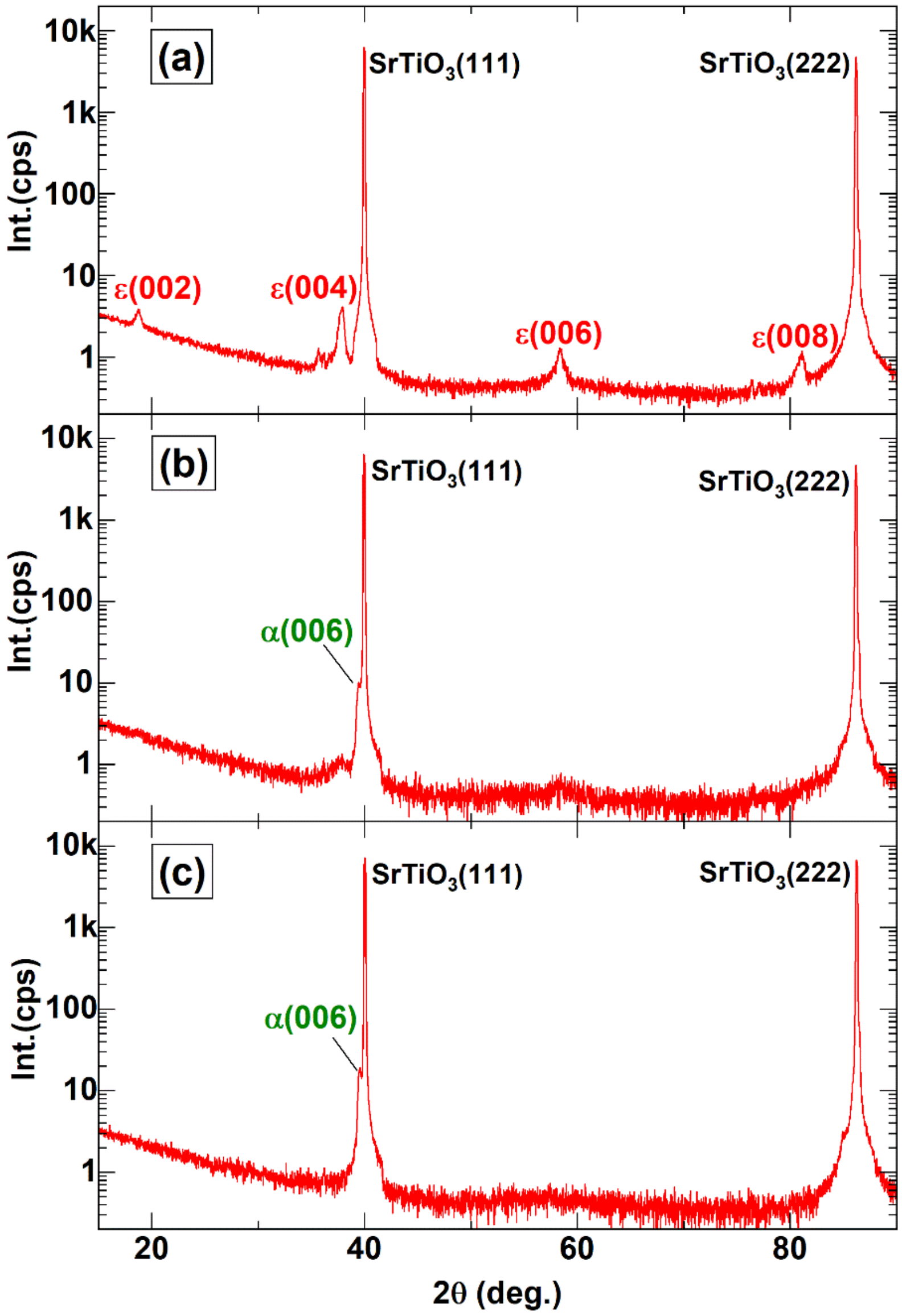

Fig. 8. θ/2θ scan XRD profiles for the α- and ε-$Fe_2O_3$ epifilms grown on $SrTiO_3$ (111) with sputtering Ar+25%$O_2$ mixed gas pressure P(Ar+25%$O_2$) = (a) 1 mTorr, (b) 2 mTorr and (c) 5 mTorr.

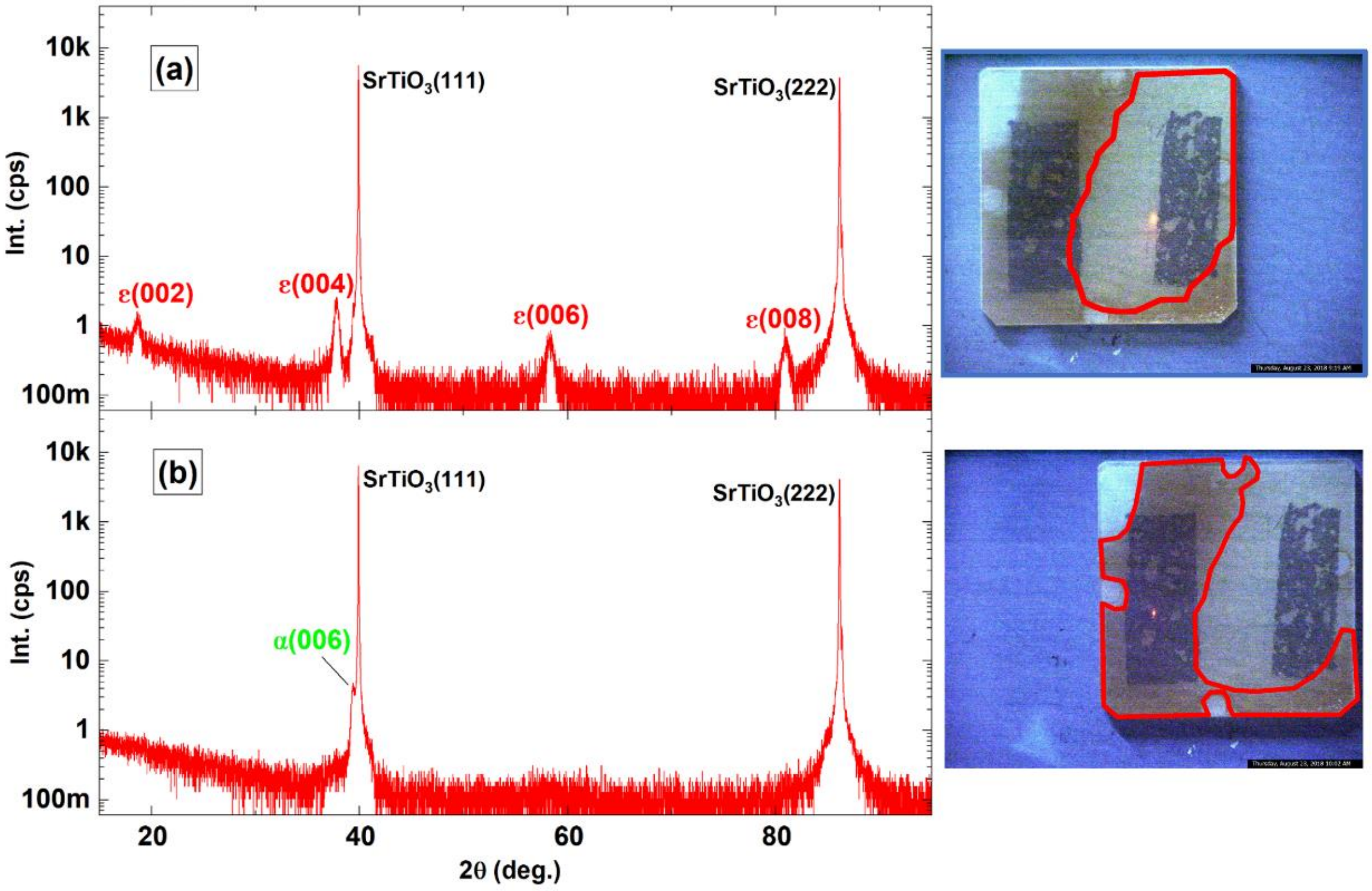

Fig. 9. θ/2θ scan XRD profiles for the α- and ε-$Fe_2O_3$ / $SrTiO_3$ (111) epifilm (P(Ar+40%$O_2$) = 1mTorr) in (a) the light-colored area (4nm thick) and (b) the dark-colored area (29nm thick) in the right side sample images (areas surrounded by red lines).

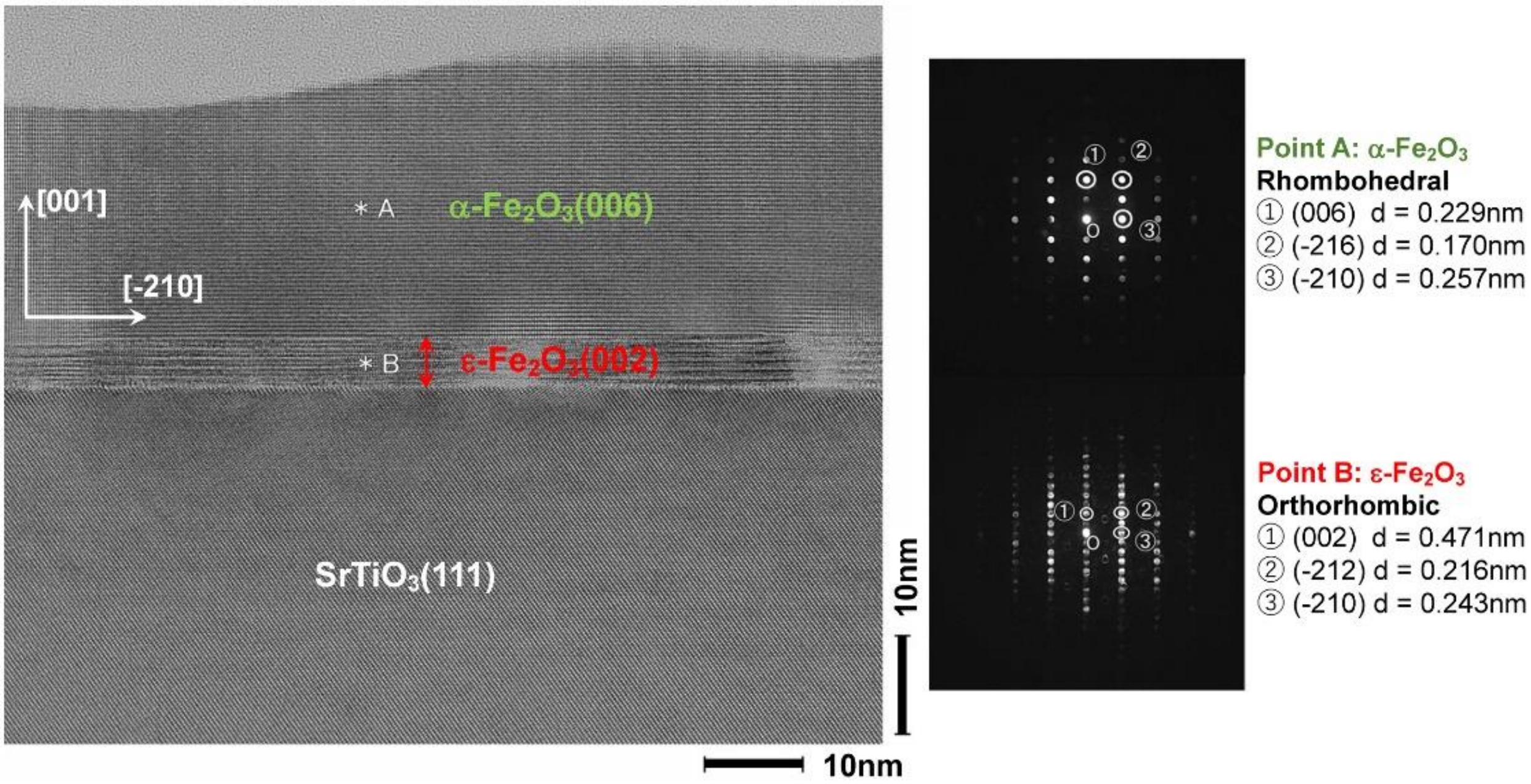

Fig. 10. Cross-sectional TEM image and nano beam electron diffraction (NBD, point A and B in the TEM image) for the α- and ε-$Fe_2O_3$ / $SrTiO_3$ (111) epifilm (P(Ar+40%$O_2$) = 1mTorr) in the dark-colored area shown in Fig. 9.

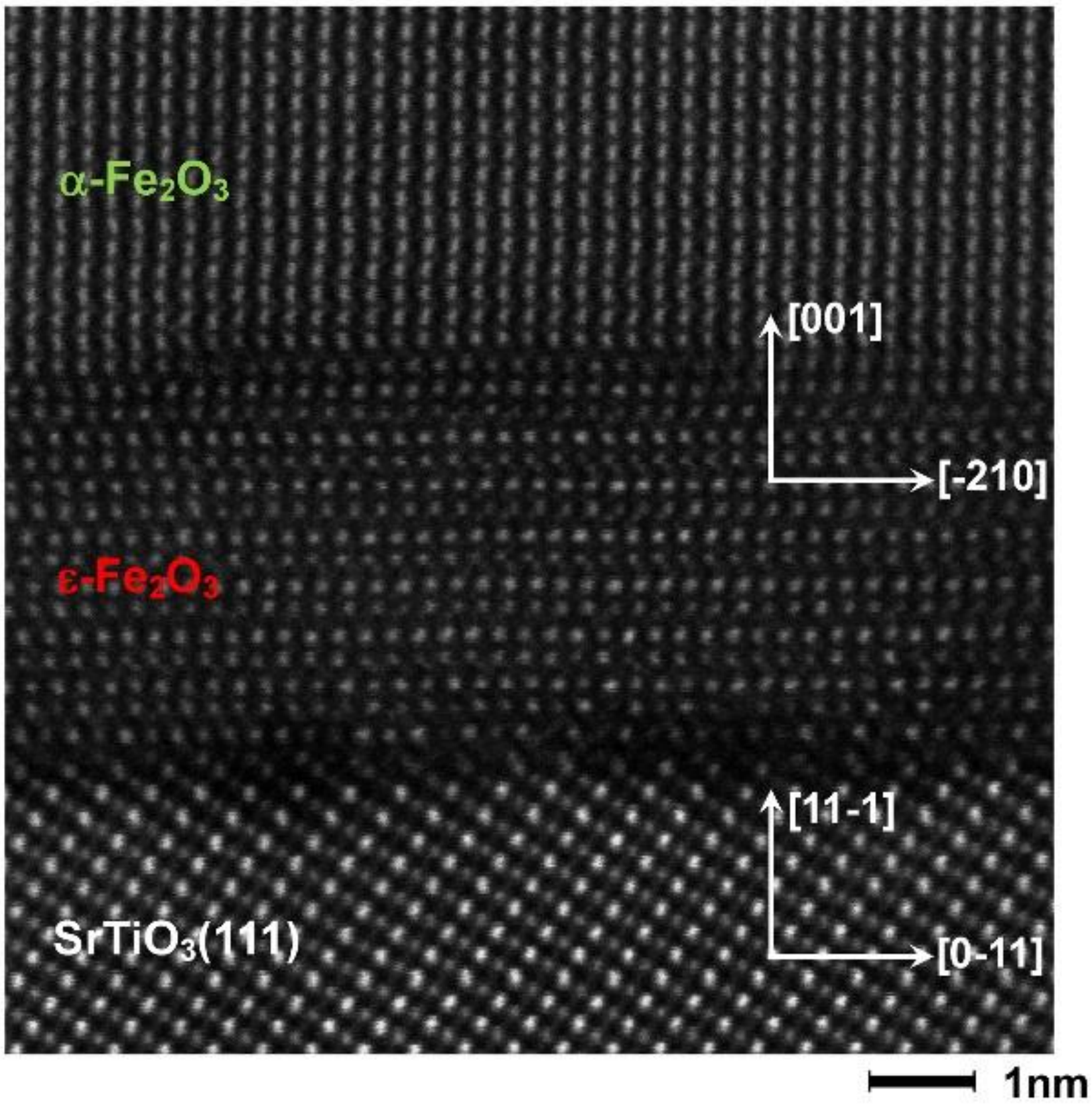

Fig. 11. Cross-sectional HAADF-STEM (Z-contrast) image for the α/ε-$Fe_2O_3$ / $SrTiO_3$ (111) epifilm (P(Ar+40%$O_2$) = 1mTorr) in the dark-colored area shown in Fig. 9.

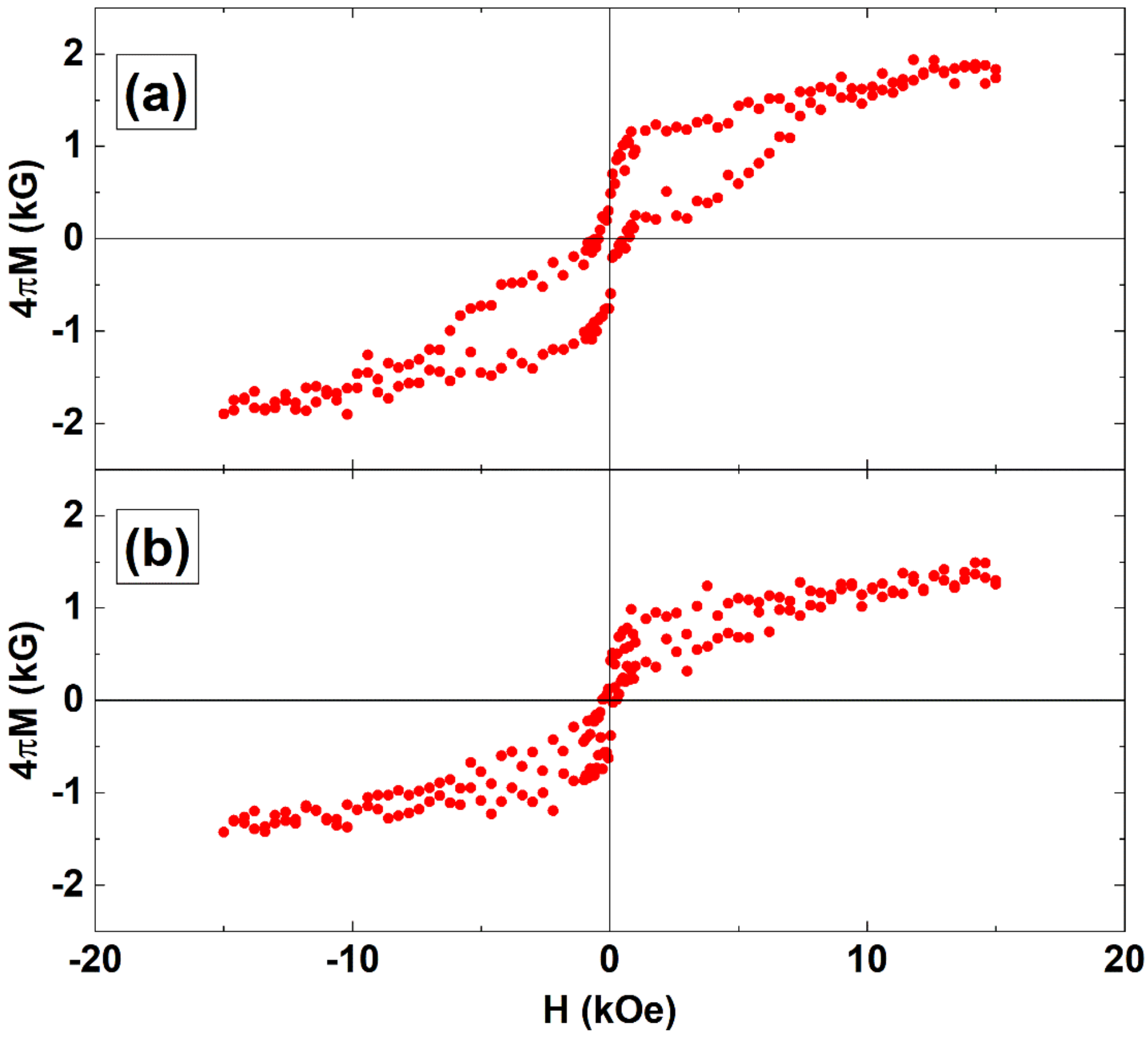

Fig. 12. Room-temperature in-plane magnetization curves for the α- and ε-$Fe_2O_3$ / $SrTiO_3$ (111) epifilm (P(Ar+40%$O_2$) = 1mTorr) with total thicknesses = (a) 121 nm and (b) 29 nm. Magnetizations 4πM were evaluated assuming 5nm thickness of the ε-$Fe_2O_3$ initial epilayer.

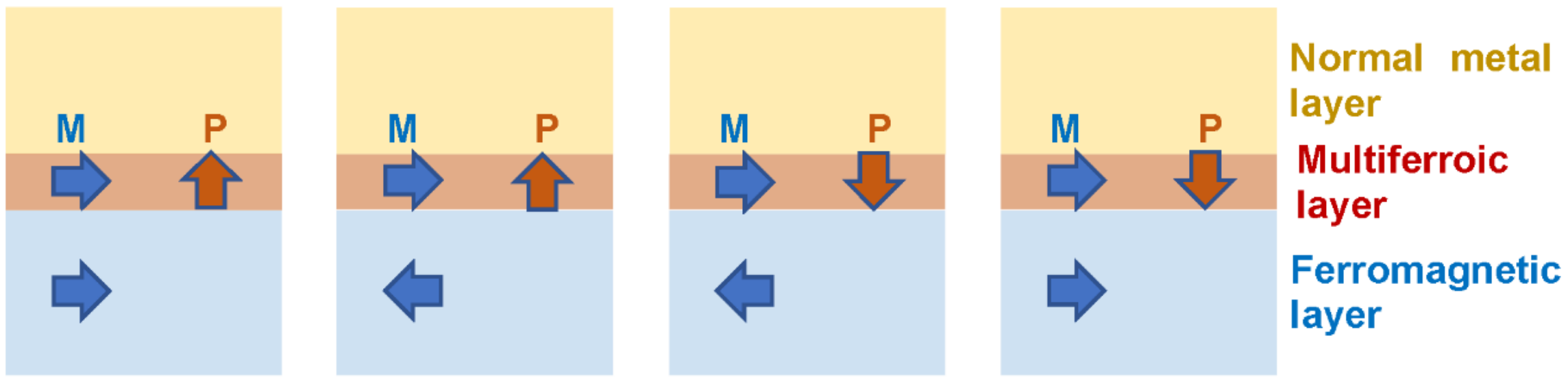

Fig.13. Schema of 4-resistive-state multiferroic tunnel junction (MFTJ) utilizing the spin filter effect in multiferroic barrier such as ε-$Fe_2O_3$ thin layer and tunnel electroresistance (TER) effect. M and P represent magnetic moment and electric polarization, respectively.